%% file: ms.tex
\documentclass{article}

\usepackage{graphicx}
\usepackage{listings}
\usepackage{enumitem}
\usepackage{hyperref}
\usepackage{algorithm}
\usepackage{algorithmic}
\usepackage{float}
\usepackage{bm, bbm}
\usepackage{subfig}
\usepackage{amsmath,amssymb,mathrsfs}
\usepackage{url}
\usepackage{wrapfig}
\usepackage{multirow}
\usepackage[normalem]{ulem}
\usepackage{booktabs}

\usepackage{etoolbox}
\usepackage{url}
\usepackage[most]{tcolorbox}
\usepackage{xcolor}
\usepackage{xspace}
\usepackage{subfig}

\include{macro_wnchen}

\newcommand{\cpk}{$\msf{CP}$-$\kappa$\xspace}
\newcommand{\cpd}{$\msf{CP}$-$\Delta$\xspace}

\usepackage[round]{natbib}   
\newtheorem{theorem}{Theorem}[section]

\newtheorem{lemma}[theorem]{Lemma}

\newtheorem{definition}[theorem]{Definition}

\title{Randomization Techniques to Mitigate the Risk of Copyright Infringement}
\author{Wei-Ning Chen$^1$\thanks{Work done during Wei-Ning's internship at Google.} \and Peter Kairouz$^2$ \and Sewoong Oh$^{2,3}$ \and Zheng Xu$^2$ } 
\date{
Stanford University$^1$ \,\, Google$^2$ \,\, University of Washington$^{3}$
} 

\begin{document}
\maketitle
\begin{abstract}
    In this paper, we investigate potential randomization approaches that can complement current practices of input-based methods (such as licensing data and prompt filtering) and output-based methods (such as recitation checker, license checker, and model-based similarity score) for copyright protection. This is motivated by the inherent ambiguity of the rules that determine substantial similarity in copyright precedents. Given that there is no quantifiable measure of substantial similarity that is agreed upon, complementary approaches can potentially further decrease liability. Similar randomized approaches, such as differential privacy, have been successful in mitigating privacy risks. This document focuses on the technical and research perspective on mitigating copyright violation and hence is not confidential. After investigating potential solutions and running numerical experiments, we concluded that using the notion of Near Access-Freeness (NAF) to measure the degree of substantial similarity is challenging, and the standard approach of training a Differentially Private (DP) model costs significantly when used to ensure NAF. Alternative approaches, such as retrieval models, might provide a more controllable scheme for mitigating substantial similarity.
\end{abstract}

\input{sec1_introduction}
\input{sec2_prelimenary}

\input{sec3_our_approach}

\newpage
\bibliography{ms}
\newpage
\input{sec99_appendix}

\end{document}

%% file: macro_wnchen.tex
\newcommand{\lp}{\left(}
\newcommand{\rp}{\right)}

\newcommand{\lba}{\left\lvert}
\newcommand{\rba}{\right\rvert}

\newcommand{\mcal}{\mathcal}

\newcommand{\msf}{\mathsf}

\newcommand{\ra}{\rightarrow}

\newcommand{\eqDef}{\triangleq}

%% file: sec1_introduction.tex
\section{Introduction}
Modern machine learning relies heavily on large amounts of high-quality training data, primarily obtained from the Internet. Inevitably, these large-scale datasets contain some copyrighted material. When models are trained on this copyrighted data, they can accidentally generate outputs that closely resemble the training data, leading to potential copyright infringement. For example, despite recent advancements in foundational models \citep{bommasani2021opportunities}, studies have shown that these models can easily memorize substantial portions of their training data \citep{carlini2021extracting, carlini2023extracting}.

This immediately leads to the following questions: How do we define copyright infringement for models trained on potentially copyrighted data? How can one claim that a model violates copyright laws? And how can we prevent the models from generating outputs that resemble significantly copyrighted data? Copyright laws aim to promote creativity while protecting the rights, often economically, of original works. Note that copyright infringement class-action cases can be lucrative, which accounts for their popularity compared to, for instance, privacy violation cases. Under the fair use doctrine \citep{CopyrightLaw2022}, reproducing copyrighted work can be considered fair use based on four factors: the purpose of the use, the nature of the work, the amount of similarity, and potential harm. While purpose, nature, and harm are outside the scope of engineering solutions (see, for example, \citet{sag2018new, sobel2017artificial} for a discussion on whether data mining and machine learning on copyrighted text falls under ``fair use''), substantial similarity can potentially be addressed with technology. 

Producing outputs substantially similar to copyrighted work on which a foundation model is trained can be a key factor in determining whether it constitutes fair use. Current ongoing
lawsuits that do not demonstrate that foundation models generate substantially similar works are likely to be dismissed. Determining the amount and substantiality of the portion of the generated text in relation to the copyrighted work is subjective and ambiguous as courts look at both quantity and quality. Fair use is less likely to be found if the use includes a large portion of the copyrighted work. However, some courts have found the use of an entire work to be fair under certain circumstances. In other contexts, using even a small amount of a copyrighted work was determined not to be fair because the selection was an important part, or the ``heart'', of the work, e.g., Ford vs. Nation magazine. Despite such challenges, there have been attempts to come up with quantifiable metrics of substantial similarity and corresponding guidelines.


\paragraph{Measuring substantial similarity.}
Suppose we have an oracle such that when presented with an original work $x$ and an allegedly-infringing work $y$, outputs a binary decision $\msf{sim}(x,y)$ on whether they are substantially similar or not. This naturally leads to the following output filtering approach (e.g., \citet{xu2021detoxifying})to copyright protection: After generating a text output $y$, one enumerates all copyrighted works to check for substantial similarity. Variations of output filtering are present in most large language model services for various reasons, including copyright, safety, alignment, etc. 

\paragraph{Divergence-based metric.}
Moving away from the typical output filtering framework, \citet{scheffler2022formalizing} introduced a new framework in the context of comparing two computer programs in an attempt to make the notion of substantial similarity formal. The main idea is to use the minimum description length of a program to generate the allegedly-infringing work $y$, with and without access to the original work $x$. If the description lengths do not differ more than some threshold, then the contribution of $x$ in generating $y$ is small, and one can assert that the work has enough novelty and merits fair use. Although minimum description lengths are not easy to compute, the idea of comparing two scenarios with and without access to the original work was a major step forward in formalizing substantial similarity, which led to several important new approaches. This has obvious parallels with differential privacy, where paired datasets are tested on whether a piece of sensitive information was included or not. This connection between copyright and DP was initially suggested in \citet{bousquet2020synthetic}. However, there are challenges that are commonly shared among the work that follows this approach of paired scenarios.

\begin{itemize}
    \item \textbf{Difficulty in specifying the unit of original work}: When we compare the current trained model against one that leaves a piece of copyrighted work out, the designer is left to choose how much of a work to leave out. Should it be one volume in the Harry Potter series, the entire series, or a chapter in a book? Since the expressions, and not the styles or ideas, are copyrighted, \citet{vyas2023provable} argues that smaller units that qualify as an isolated piece of expression should suffice. For example, a painting as opposed to the entire collection of an artist. This choice is important but will likely evolve over time, and algorithmic solutions should be flexible to changes in the choice of the unit of copyright. For now, it can be treated as a hyper parameter to be chosen at the training time.

    \item \textbf{Derivative works:} A related concern is that internet-scale data is interconnected. Excerpts, quotes, and fan fictions are abundant and it is impossible to completely isolate all derivative work of an original work, which most approaches in this direction require.
\end{itemize}

Some preprocessing to identify the boundaries of each original work in an internet-scale dataset is necessary for copyright protection techniques, which might be orthogonal to the technical solutions investigated in this paper as long as they are flexible to changing boundaries.

In this paper, we first overview recently proposed techniques to mitigate the production of substantially similar outputs, examine their merits and weaknesses, and then evaluate and enhance these techniques by leveraging the potential for randomization in modern language models.

%% file: sec2_prelimenary.tex
\section{Prelimenary on Near Access-Freeness}
A crucial property of typical modern generative models is that the outputs are generated randomly. There are various techniques to sample the outputs, but most of them involve some randomness. Leveraging this randomness, \citet{vyas2023provable} proposed Near Access-Freeness (NAF) as a quantifiable metric for determining substantial similarity. This hinges on the inherent randomness in modern generative models, where the output is sampled from some distribution. This is based on the divergence between the output distribution of the potentially infringing language model and a \emph{safe} model that does not have access to the original work in question.

Specifically, \cite{vyas2023provable} use the abstraction of a function safe that maps a datapoint $C \in \mcal{C}$ into a generative model $\msf{safe}(C) \in \mcal{M}$ that is assumed to have been trained without any access to $C$. For example, the leave-one-out-safe function is one such example. In this construction, the safe model is trained on all data except for $C$.

Since $\msf{safe}(C)$ is a generative model that was learned without access to $C$, in many realistic scenarios, the probability that $\msf{safe}_C(·|x)$ generates material that is similar to $C$ itself will be exponentially small in the length of $C$. Moreover, even if this unlikely event happened, this generation can be said to be fortuitous.

Formally, \cite{vyas2023provable} defines the following notion of NAF:
\begin{definition}\label{def:naf}
    Let $\mcal{C}$ be a set of copyrighted data points (i.e., samples) and $\mcal{M}$ be a collection of (trained) models. Let $\msf{safe}: C \ra \mcal{M}$; and let $\Delta$ be a divergence measure between distributions. We say that a generative model $p$ is $k_x$-near access-free ($k_x$-NAF) on prompt $x \in \mcal{X}$ with respect to $\mcal{C}$, $\msf{safe}$, and $\Delta$ if for every $C \in C$,
    $$ \Delta\lp p(\cdot|x) \middle\Vert \msf{safe}_C(\cdot|x) \rp \leq k_x. $$
\end{definition}

If a model $p(\cdot |x)$ satisfies NAF with $k_x = 0$, then it is exactly the same as a safe model. Any generation of text that is substantially similar to a copyrighted work is by chance, and this chance is the same as a model that has never seen the original work. Therefore, it is safe to claim that the model $p(\cdot |x)$ is not infringing copyright. Generally, if a generative model satisfies NAF with a small $k_x$, then one can claim it is less likely that the model is outputting something substantially similar with a probability that is much larger than a random chance.

\subsection{Achieving Near Access-Freness}\label{sec:cpd_and_cpk}
In \citet{vyas2023provable}, two algorithms, \cpd and \cpk, are provided to achieve provable NAF guarantees, which we briefly overview in the following.

The \cpd algorithm (Algorithm~\ref{alg:cpd}) can be viewed as a model ensemble method, which combines multiple models trained on different partition of the training data and is used to protect from copyright infringement. 

\begin{algorithm}[hb]
  \caption{CP-$\Delta$: Copy Protection w.r.t. divergence $\Delta$ (Algorithm~3 of \citet{vyas2023provable}}\label{alg:cpd}
  
  \algorithmicrequire{ Dataset $\mcal{D}$, and divergence $\Delta \in \{\Delta_{\text{max}}, \Delta_{\text{KL}}\}$.}

  \begin{algorithmic}[t]
    \STATE Partition $\mcal{D}$ into two disjoint sets $\mcal{D} = \mcal{D}_1 \cup \mcal{D}_2$ (ideally with similar size).
    \STATE Train two safe generative models $\mcal{M}_1 = q_1(\cdot |x)$ and $\mcal{M}_2 = q_2(\cdot|x)$ with $\mcal{D}_1$ and $\mcal{D}_2$, respectively, where $x$ is the prompt to the model.
    \STATE For any given prompt $x$, generate sample $y$ according to
    $$ p(y|x) = 
    \begin{cases}
    \frac{\min\{q_1(y|x), q_2(y|x)\}}{Z(x)};\\
    \frac{\sqrt{q_1(y|x)\cdot q_2(y|x)}}{Z(x)},
    \end{cases}$$
    where $Z(x)$ is the partition function (i.e., the normalization constant).
  \end{algorithmic}
\end{algorithm}

While \cpd is proven to be $k_x$-NAF for $k_x \leq -\log\lp 1-\msf{TV}\lp q_1(\cdot |x), q_2(\cdot | x) \rp \rp$ under $\Delta_\msf{max}$ and $k_x \leq -2\log\lp 1-\msf{H}^2\lp q_1(\cdot |x), q_2(\cdot | x) \rp \rp$ under $\Delta_\msf{KL}$ \citep[Theorem~3.1]{vyas2023provable}, there are several drawbacks that make \cpd infeasible in many practical deployments. First, implementing \cpd may be computationally difficult, especially if $q_1$ and $q_2$ are autoregressive models for text sequences or diffusion models for image generation\footnote{Note that, however, one can implement \cpd with rejection sampling and partially circumvent the computational issue. See Section~\ref{sec:cpd_rejected} for details.}. Second, the bound on $k_x$ is a model-dependent quantity that is hard to calculate or estimate in practice. Therefore, a more flexible algorithm, \cpk, is proposed.

\begin{algorithm}[h]
  \caption{CP-$\kappa$: Access-Free Reduction at Threshold $\kappa$ (Algorithm~4 of \citet{vyas2023provable}}\label{alg:cpk}
  \algorithmicrequire{ a model $p$, a set of safe models $\msf{safe}$, a threshold $\kappa$.}
  \begin{algorithmic}[t]
    \STATE Sample $y \sim p(\cdot |x)$ and return $y$ if
    $$ \forall q \in \msf{safe},\, \log(p(y|x) / q(y|x)) \leq \kappa. $$
  \end{algorithmic}
\end{algorithm}

By introducing a tunable parameter $\kappa$ and adopting rejection sampling, \cpk resolves the computation issue in generating a sample, provided the threshold $\kappa$ is sufficiently high. In addition, under $\Delta_\msf{max}$, \cpk achieves the $k_x$-NAF for any $k_x > \kappa + \log(1/\nu_\kappa(x))$, where $\nu_\kappa(x)$ is the probability that a sampled $y$ is accepted in a single iteration of the while loop in Algorithm~\ref{alg:cpk} \citep[Theorem~3.5]{vyas2023provable}. Although $\nu_\kappa(x)$ cannot be computed exactly, it can be accurately estimated by repeating the algorithm assuming $\kappa$ is not too large. his approach provides a method to obtain a probabilistic upper bound on the model-dependent NAF guarantee $k_x$.

\subsection{Connection to differential privacy}
There are obvious connections between NAF and DP \citep{elkin2023can}. Different choice of $\Delta$ corresponds to different variants of DP (e.g., the standard $\varepsilon$-DP in \citet{dwork2006calibrating} when $\Delta = \Delta_\msf{max}$ or $(1, \varepsilon)$-R\'enyi DP in \citet{mironov2012significance} when $\Delta = \Delta_{\msf{KL}}$). Adopting the original notion of DP to the context of generative models, we get the following definition:
\begin{definition}[Differentially private generation]
    For two neighboring datasets $S$ and $S'$, let $P_{S}(\cdot|x)$ denote the probability distribution of a generated text on input prompt $x$ when the generative models are trained on a dataset $S$ with an algorithm $\mcal{A}$, where the randomness includes the internal randomness in the training algorithm $\mcal{A}$ and the randomness in the generation. We say this generation of a single output is an $\varepsilon$-Differentially Private Generation ($\varepsilon$-DPG) if for every neighboring dataset $S$ and $S'$, every prompt $x \in \mcal{X}$, it holds that
    $$ \Delta\lp P_{S}(\cdot|x) \Vert P_{S'}(\cdot|x) \rp \leq \varepsilon, $$
    where $\Delta$ is some divergence (e.g., $\Delta_\msf{max}$ or $\Delta_\msf{KL}$).
\end{definition}

Recall that two datasets $S$ and $S'$ are called neighbors if they only differ in one unit of privacy, which is typically one sample but could be larger depending on the context. If a single generative model is trained with $\varepsilon$-DP (i.e., the standard DP applied to the \emph{trained model}), then any generated text satisfies $\varepsilon$-DPG by the data-processing inequality. The advantage $\varepsilon$-DPG is that it allows one the flexibility to add randomness at the generation stage rather than the training stage, which can potentially give a significant gain in the utility-privacy tradeoff. However, there are a few major differences between the standard $\varepsilon$-DP and $\varepsilon$-DPG. First, for $\varepsilon$-DPG, multiple generated texts can eventually reveal any private data in training, whereas $\varepsilon$-DP protects against any number of generations. Next, the generative model can be shared under $\varepsilon$-DP, whereas only the generated text can be shared under $\varepsilon$-DPG.

\citet{elkin2023can} points out a few differences between NAF and DPG. First, NAF is one-sided whereas DPG is symmetric. This can give some flexibility in designing algorithms that achieve better utility under the one-sided NAF. Next, NAF allows more flexibility in selecting what safe model to use in the definition. Given the similarity between the definitions of NAF and DPG, it is natural to build upon this connection and consider using DP or DPG algorithms to achieve NAF. In this paper, we propose to leverage tools from DP to address issues in previous methods that guarantee NAF. 

\subsection{Challenges in previous solutions and summary of our contributions} \label{sec:naf_challenges}
Although the solutions proposed in \citet{vyas2023provable}, such as \cpk and \cpd (see Section~\ref{sec:cpd_and_cpk}), have resolved some of the computation issues, there are still challenges that need to be further addressed:
\begin{itemize}
    \item \textit{Computational challenges for verifying NAF:} Even for a given instance of $(\msf{C}, \msf{safe}, \Delta, x)$ and an arbitrary generative model $p(\cdot|x)$, the computational cost for checking the NAF condition generally scales with the support of the output of the language model. This is astronomical for the large language models we are interested in. Furthermore, the autoregressive nature of the decoding process makes this even more challenging. Approximating it with a truncated support is problematic, as typical choices of divergences (such as $\Delta_\msf{KL}$ or $\Delta_\msf{max}$) are sensitive to the tail of the distribution. This makes it difficult to compare two generative models with respect to their respective achieved NAF, even if a reference safe model and a prompt $x$ are given.

    While efficient estimation of $k_x$ is possible in some limited scenarios (such as \cpk), in general, the estimation scheme does not extend to a general model $p(\cdot |x)$. Note that in \cpk, additional rejection sampling is required, so the final model, so the final model $p_\kappa(\cdot|x)$ should be treated as an ensemble of $p$ and all the safe models. Since the NAF guarantee $k_x$ is a data- and model-dependent quantity (i.e., it depends on the prompt $x$, the safe models, and the reference model $p(\cdot|x)$, and should be denoted as $k_x\lp p; \msf{safe}, x \rp$), ideally, we want to have a ``audit'' scheme that provides an empirical estimate of $\hat{k}_x$ with sufficient confidence.
    

    \item \textit{Unclear advantage over DP-based methods:} While \citet{vyas2023provable, elkin2023can} identified key differences between copyright protection and differential privacy, mathematically, DP remains a stricter criterion compared to NAF.  Specifically, $k$-(model) DP implies $k$-DPG (for any prompt $x$), which in turn implies $k$-NAF. In \citet{vyas2023provable}, the NAF guarantees of the generative models range from $10^1$ to $10^{3}$, making it unclear whether, under such a large privacy budget, DP-based methods are still strictly worse, in terms of the utility, than the proposed NAF algorithms like \cpd and \cpk. Additionally, the DP-based solution provides a \emph{worst-case} NAF guarantee independent of the prompt $x$ and safe models, which can be favorable in some scenarios. One can achieve a ``safer'' guarantee (e.g., smaller $\varepsilon$ for DP, or, effectively, smaller $k$ in NAF) by adjusting the injected noise accordingly. On the other hand, existing NAF algorithms solely rely on the stability of the safe models, so when the safe models do not align with each other (i.e., when the divergence between $q_1(\cdot|x)$ and $q_2(\cdot|x)$ is large), it may be impossible to achieve a pre-specified strict NAF guarantee.
    
    \item \textit{The case of substantially similar outputs:} NAF is defined over (the distribution of) all outputs and not just those similar to the original work $c$ of interest. This choice is unnecessarily pessimistic, and a crucial aspect of the challenge has been forgotten; fair use only concerns the substantial similarity in the expression of the output. An earlier approach \citet{scheffler2022formalizing} is deterministic and only checks for outputs similar to $c$ while still comparing two programs with and without access to the original work $c$. \citet{scheffler2022formalizing} avoid explicitly specifying what constitutes as substantially similar by using Minimum Description Lengths (MDL) and the resilience that comes with this metric. However, one pays the heavy computational cost of computing MDLs. Furthermore, there is no efficient algorithm that guarantees a desired level of the MDL-based metric.

    \item \textit{Difficulty in specifying the reference safe models:} NAF assumes a safe model is given by an external entity. In a fair use case, if NAF is to be used, the defendant will need to produce a language model that (1) did not access the original copyrighted work and (2) outputs text with distribution close to the allegedly infringing language model. The fact that the defendant can choose different safe models for each prompt $x$ and each original work $c$ makes the notion of NAF unreliable. One with more resources could come up with better, safer models and claim a smaller NAF. This brittleness in the definition safe model leaves NAF open to criticism. On the other hand, there are algorithmic approaches that ensure NAF with specific safe models that could be concrete or theoretical.
\end{itemize}

\paragraph{Our contributions.} 
In this work, we aim to address the first and second challenges by proposing improved solutions and conducting comprehensive experiments to evaluate them empirically. First, we compare the performance of \cpk (Algorithm~\ref{alg:cpk}) with a DP-based solution (trained with DP-FedAvg \citep{mcmahan2016communication}) on a next-token prediction task. Next, we evaluate how \cpk and \cpd can mitigate memorization in a fine-tuning task. As a by-product, we propose a Monte Carlo method to empirically estimate the NAF guarantee (i.e., $k_x$) for sentence-level generation. Finally, to further achieve a stricter NAF guarantee (e.g., stricter than the $k_x$ that \cpd or \cpk provides), we propose adding additional randomization into the generation process, such as increasing the decoding temperature, performing randomized response, or interpolating with an $\varepsilon$-DP model.

%% file: sec3_our_approach.tex
\section{Methodology and Empirical Evaluations}
In this section, we conduct experiments to evaluate prior methods and propose solutions to address the challenges described in Section~\ref{sec:naf_challenges}. 

\subsection{Monte Carlo method for estimating NAF guarantees}
We start by introducing the computation and estimation of NAF guarantees, specifically $k_x$. For token-level generation, the divergences $\Delta_\msf{KL}$ and $\Delta_\msf{max}$ between two models (e.g., outputs from \cpd/\cpk and safe models) are tractable and can be computed exactly. However, for sentence-level generation, the computational cost grows exponentially fast, i.e., $O(K^T)$ where $K$ is the number of tokens and $T$ is the sentence length. 

While some methods, such as \cpk (with a sufficiently high threshold \(\kappa\)), can yield an efficient and straightforward estimate of the corresponding \(k_x\), these estimators are typically tailored to the specific algorithm and do not extend to general scenarios. Therefore, we propose using the following Monte Carlo estimator: for $\Delta = \Delta_\msf{KL}$, a model $p(\cdot|x)$ and pre-specified safe models $\msf{safe} = \{q_1(\cdot|x), q_2(\cdot|x), ..., q_m(\cdot|x)\}$,
\begin{align}\label{eq:mc}
    &\hat{k}_x\lp p; \msf{safe}, x \rp \eqDef \max_{j \in [m]} \hat{\Delta}_j, \, \text{where }
\end{align}
$$ \hat{\Delta}_j \eqDef \frac{1}{n}\sum_{i = 1}^n \log\lp \frac{p(y_i|x)}{q_j(y_i|x)} \rp, $$
and $y_1, ..., y_n$ are independent samples generated from $p(\cdot|x)$. Note that this yields a consistent estimate of the true upper bound (i.e., \(k_x\)), as each term of \(\hat{\Delta}_j\) is an unbiased estimator of the divergence. Moreover, we can reduce the variance by replacing these estimates with
$$\hat{\Delta}_j \eqDef \frac{1}{n}\sum_{i=1}^n \log\left(\frac{p(y_i|x)}{q_j(y_i|x)}\right) - \left(\frac{p(y_i|x)}{q_j(y_i|x)} - 1\right),$$
which ensures that \(\hat{\Delta}_j\) is always positive.

When $p \ll q_j$ (e.g., when $p$ is obtained from \cpd or \cpk), one can further apply the empirical Berstein inequality \citep[Theorem~3]{maurer2009empirical} to derive a high probability bound, which we state in the following lemma:
\begin{lemma} 
    Assume $p(\cdot|x) \ll q_j(\cdot|x)$ for all $j \in [m]$. If $p(y|x) \geq \alpha$ and $q_j(y|x) \geq \alpha$ for all $y \in \msf{supp}\lp q_j(\cdot|x) \rp$, then it holds that, with probability $1-\delta$
    \begin{align}
        \lba \hat{k}_x - k_x \rba \leq & \sqrt{\frac{8\max_{j\in[m]}V_n\lp r_j^n \rp\log(1/\delta)\log^2(m/\alpha)}{n}} \nonumber \\
        & + \frac{14\log(2/\delta)\log(m/\alpha)}{3(n-1)},
    \end{align}
    where $k_x \eqDef \max_j \Delta_\msf{KL}(p(\cdot|x), q_j(\cdot|x))$, $r_j^n \eqDef \lp p(y_i|x)/ q_j(y_i|x) \rp_{i \in [n]}$ and $V_n(r_j^n)$ is the sample variance of $r_j^n$.
\end{lemma}
The proof follows from a simple application of the empirical Berstein inequality, together with the fact that $\lba \log(p(y|x)/q_j(y|x)) \rba \leq 2\log(1/\alpha)$. This implies that the sample complexity is roughly $O\lp \log^2(1/\alpha) \rp$ (when other factors are constant). 

For general language models, \(\alpha\) can be arbitrarily small as there is no constraint on the generation probability. However, if one adopts popular techniques such as nucleus sampling \citep{holtzman2019curious} or top-\(p\) decoding (where we only keep the largest tokens whose cumulative probability exceeds \(p\) at each stage of decoding), an upper bound on \(\alpha\) can be obtained:

$$ \alpha \geq \left(\frac{1-p}{K}\right)^T, $$

where \(K\) is the token size and \(T\) is the sequence length. This suggests that the sample complexity of obtaining an accurate estimate is roughly \(O\left(T^2 \log(1/K)\right)\) \footnote{While the sample complexity may still seem pessimistic, it is poly-logarithmic in the support size. For example, if we want to evaluate the NAF bound on a 20-token sentence generation, 100-1000 independent samples suffice.}.

\subsection{Evaluating NAF bounds of \cpk and \cpd}
\paragraph{Comparison to (model) differential privacy.} In the first experiment, we focus on the language generation task and compare methods based on differential privacy and \cpd. To better define the ``unit'' of the text corpus, we use the federated StackOverflow dataset\footnote{Admittedly, the model is less expressive than modern transformer-based architectures, given the constraints of available resources for DP/federated training. However, this experiment allows us to effectively assess the trade-offs between NAF and utility.}, treating each user's data as a unit. We then train a 4M-parameter LSTM model (details are given in Appendix~\ref{sec:lstm_details}) on this data for the next-word prediction task with differential privacy and compare the results with those obtained using \cpd. For \cpd, we split the training data into two disjoint sets and train two safe models, respectively.
\begin{figure}[h]
    \centering
    \includegraphics[width=0.95\linewidth]{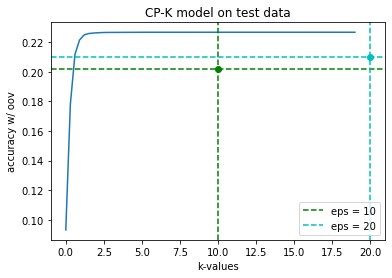}
    \caption{A comparison between $k$-NAF (with $\Delta_\msf{max}$) and $\varepsilon$-DP (with $\delta = 10^{-6}$).}
    \label{fig:naf_dp}
\end{figure}

\paragraph{NAF of \cpk and \cpd on Federated Stackoverflow.} Next, we visualize the $k_x$ values of \cpd and \cpk (with different thresholds $\kappa$) on the test data. Note that the $k_x$ values here correspond to the divergence of the next-token prediction. We plot the histograms of $k_x$ with different $x$'s.

\begin{table*}[h]
    \centering
    \begin{tabular}{c|c|c|c|c|c|c}
& Loss test & Acc test & Loss train 1 & Acc train 1 & Loss train 2 & Acc train 2 \\
\hline \hline
Safe Model 1 & 3.4599    & 0.227    & 3.1235       & 0.256       & 2.7554       & 0.267       \\
Safe Model 2 & 3.5625    & 0.218    & 3.2495       & 0.244       & 2.7999       & 0.264       \\
All (unsafe) & 3.4537    & 0.222    & 3.1159       & 0.257       & 2.7205       & 0.272       \\
\cpd   & 3.4987    & 0.226    & 3.1804       & 0.253       & 2.7653       & 0.268    
    \end{tabular}
    \caption{Losses of safe models and \cpd models on training and testing datasets.}
    \label{tab:my_label}
\end{table*}

\begin{figure}[h]
  	\centering
  	\subfloat{{\includegraphics[width=0.45\linewidth]{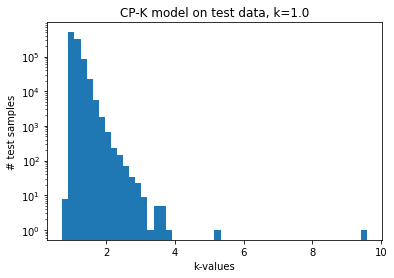} }}%
  	\subfloat{{\includegraphics[width=0.45\linewidth]{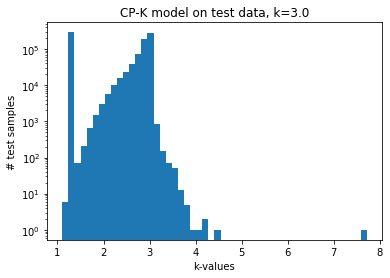} }}%
   \\
        \subfloat{{\includegraphics[width=0.45\linewidth]{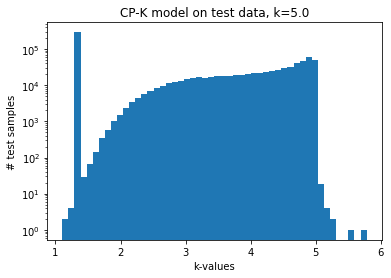} }}%
        \subfloat{{\includegraphics[width=0.45\linewidth]{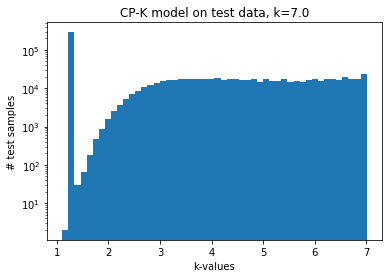} }}
  	\caption{ \cpk with threshold $\kappa = \{1.0, 3.0, 5.0, 7.0\}$ We see that decreasing $\kappa$ does not necessarily reduce $k_x$. For instance, when $\kappa = 1.0$, most $k_x$'s are still greater than $1.0$.}
  	\label{fig:cpk-thresholds}
\end{figure}

\begin{figure}[h]
    \centering
    \includegraphics[width=0.5\linewidth]{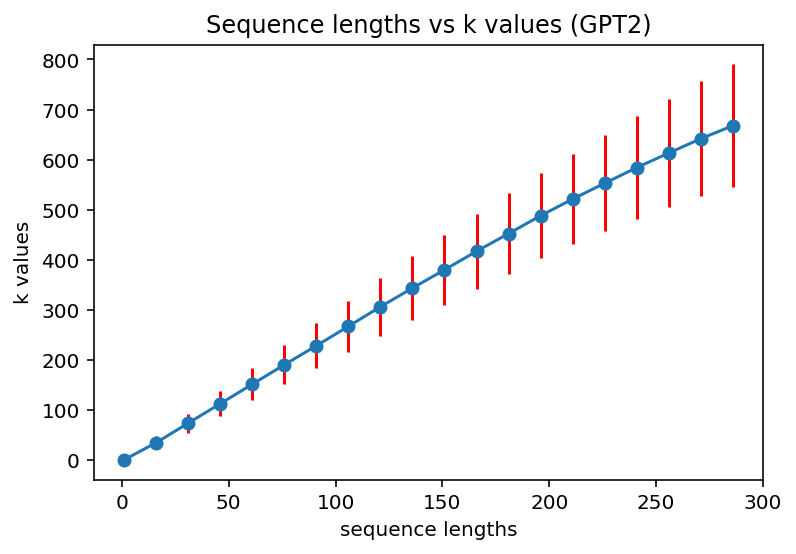}
    \caption{Sentence-level NAF bounds based on Marte Carlo estimators on a fintuned GPT-2. We use a token-level \cpd to ensemble models and plot the NAF with the sequence length.}
    \label{fig:sentence-level}
\end{figure}

\paragraph{Sentence-level NAF for \cpd.} In the next set of experiments, we evaluate the NAF bounds on a GPT-2 model fine-tuned on a PubMed dataset \citep{dernoncourt2017pubmed}\footnote{Note that the fine-tuned dataset is released after August 2022, later than the pre-trained GPT-2 model.}. We perform the Monte Carlo simulation to estimate the NAF bounds. Note that due to resource constraints, we did not perform re-sampling; instead, we calculated the empirical divergence (i.e., $\hat{\Delta}_j$ defined in \eqref{eq:mc}) and plotted the average $\hat{k}_x$ for different generation length. In Figure~\ref{fig:sentence-level}, we see that the bound scales are roughly linear with the length of the generated sequences.

\subsection{Evaluating memorization for NAF}
\begin{figure}[H]
  	\centering
  	\subfloat{{\includegraphics[width=0.49\linewidth]{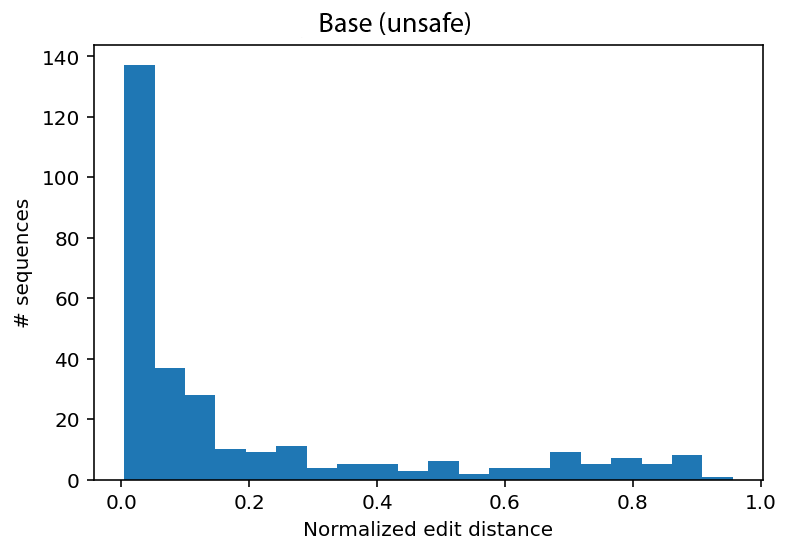} }}%
  	\subfloat{{\includegraphics[width=0.49\linewidth]{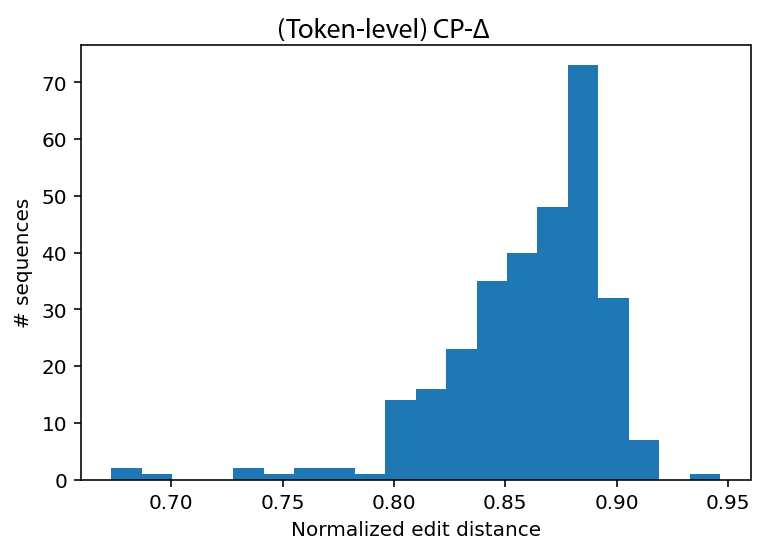} }}%
  	\caption{ Normalized edit distance between the generated samples and the true samples (duplicated in the training phase).}
  	\label{fig:memorization}
\end{figure}

Next, we empirically show that \cpd and \cpk can effectively mitigate memorization. In order to better demonstrate, we duplicate 600 training samples 40 times and trained a based model $p$ (trained on all data without any protection) and two safe models (trained on half of the training data with duplication) $q_1$ and $q_2$. To measure memorization, we feed the initial 10 tokens of the duplicated samples and compute the normalized edit distances of the outputs to the correct answers. In Figure~\ref{fig:memorization}, we plot the histogram of distances with and without protection. Based on the plot, we see that token-level \cpd can effectively mitigate memorization for sequence generation.

\subsection{Enhancing NAF via explicit randomization}
Finally, we demonstrate that a "safer" guarantee can potentially be achieved by injecting additional randomization into the decoding process. Although the \cpk algorithm includes a threshold parameter \(\kappa\), its NAF guarantee is ultimately determined by the inherent distance between the safe models \(q_1(\cdot|x)\) and \(q_2(\cdot|x)\). As shown in Figure~\ref{fig:cpk-thresholds}, even if \(\kappa\) is set to a smaller value, the NAF bound \(k_x\) cannot be reduced arbitrarily. Therefore, to achieve a better bound, we need to increase the model's randomness by either injecting more noise or making the model less certain.

\begin{figure}[h]
    \centering
    \includegraphics[width=0.95\linewidth]{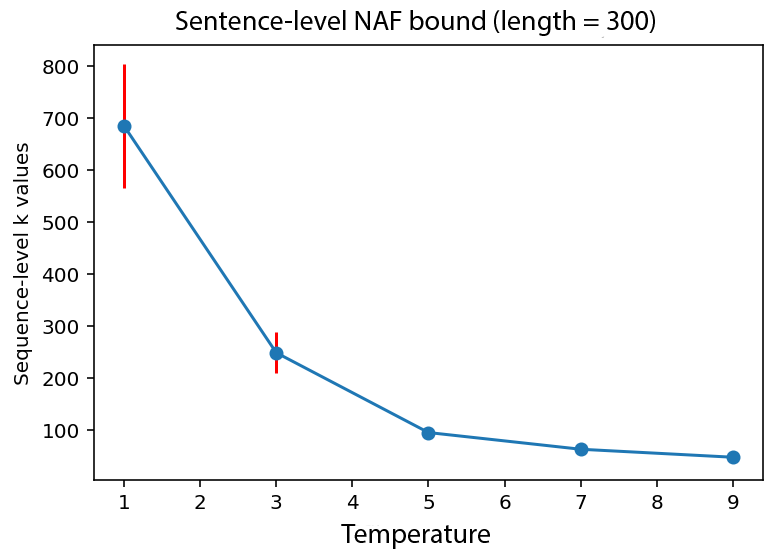}
    \caption{Sentence-level NAF bounds based with different decoding temperatures.}
    \label{fig:temperature}
\end{figure}

One simple approach to achieving this is by increasing the temperature during decoding. In Figure~\ref{fig:temperature}, we increase the temperature and plot the estimated sentence-level NAF. The results indicate that by appropriately increasing the temperature, we can obtain a better NAF guarantee.

\section{Conclusion}
In this work, we propose an alternative method based on Monte Carlo simulation to evaluate the empirical NAF guarantees. We compare the performance of the \cpk and \cpd algorithms and demonstrate how they can mitigate memorization in a fine-tuning task. To achieve a stricter NAF guarantee (e.g., stricter than the $k_x$ provided by \cpd or \cpk), we suggest incorporating additional randomization into the generation process, such as increasing the decoding temperature or performing a randomized response. Another promising direction for enhancing performance is to interpolate the outputs of a (possibly unsafe) model with an $\varepsilon$-DP model, which we plan to explore in future work.

%% file: sec99_appendix.tex
\subsection{An improved \cpd for sequence generation}\label{sec:cpd_rejected}
\begin{algorithm}[h]
  \caption{\cpd via rejection sampling for sequence generation}\label{alg:cpd_reject}
  
  \algorithmicrequire{Dataset $\mcal{D}$, and divergence $\Delta \in \{\Delta_{\text{max}}, \Delta_{\text{KL}}\}$.}

  \begin{algorithmic}[t]
    \STATE Partition $\mcal{D}$ into two disjoint sets $\mcal{D} = \mcal{D}_1 \cup \mcal{D}_2$ and train two safe models $q_1(\cdot |x)$ and $q_2(\cdot|x)$, respectively.
    \STATE Set $i \in \{1, 2\}$ uniformly at random. Also denote $i' \in \{1, 2\}, \, i' \neq i$.
    \STATE For any given prompt $x$, generate sample $y$ according to the following rule:
    \WHILE{True}
    \STATE Generate $y \sim q_i(\cdot|x)$.
    \IF {$\log(q_i(y|x)/ q_{i'}(y|x))\leq \kappa$ and $\Delta = \Delta_\msf{max}$}
    \STATE Return $y$.
    \ENDIF
    \IF {$\log(q_i(y|x)/ q_{i'}(y|x))\geq 1$ and $\Delta = \Delta_\msf{KL}$}
    \STATE Return $y$ with probability $\min\lp1, \sqrt{\frac{e^\kappa\cdot q_{i'}(y|x)}{q_{i}(y|x)}}\rp$.
    \ENDIF
    \ENDWHILE
  \end{algorithmic}
\end{algorithm}
Note that Algorithm~\ref{alg:cpd_reject} is similar to \cpk algorithm but with $p(\cdot |x)$ being set to the mixture distribution of safe models. When $\kappa = 0$, Algorithm~\ref{alg:cpd_reject} precisely recovers \cpd; however, when the safe models do not align (i.e., the divergence between $q_1(\cdot|x)$ and $q_2(\cdot|x)$ is large), the re-sampling step may be the bottleneck. 

\subsection{Model Architecture of the $4$M LSTM}\label{sec:lstm_details}
\begin{figure}[t]
    \centering
    \includegraphics[width=\textwidth]{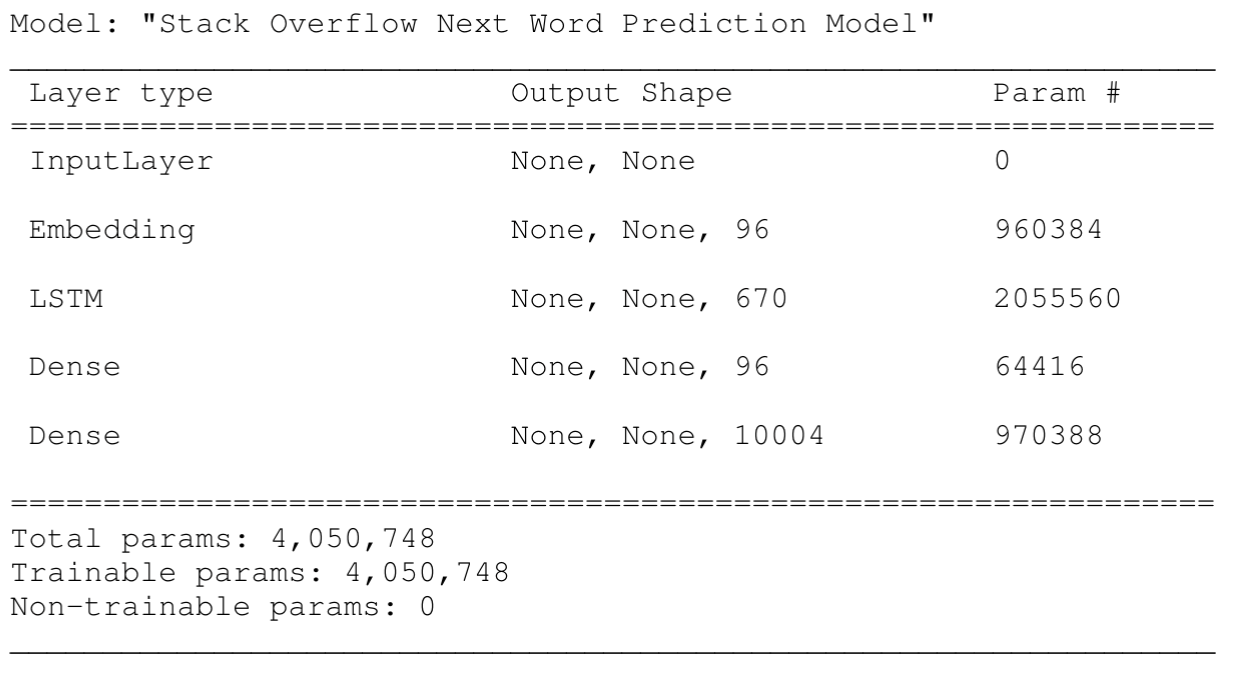}
    \caption{Stack Overflow Next Word Prediction model architecture.}
    \label{fig:sonwp-model}
\end{figure}

The model architecture is presented in Figure~\ref{fig:sonwp-model}.
